# Direct visualization of the charge transfer in Graphene/α-RuCl₃ heterostructure.

*Antonio Rossi[1,2]\*, Riccardo Dettori[3], Cameron Johnson[2], Jesse Balgley[4], John C. Thomas[2], Luca Francaviglia[2], Andreas K. Schmid[2], Kenji Watanabe[5], Takashi Taniguchi[6], Matthew Cothrine[7], David G. Mandrus[7], Chris Jozwiak[1], Aaron Bostwick[1], Erik A. Henriksen[4]\*, Alexander Weber-Bargioni[2] and Eli Rotenberg[1]*

1-The Molecular Foundry, Lawrence Berkeley National Laboratory, Berkeley, USA.

2-Advanced Light Source, Lawrence Berkeley National Laboratory, Berkeley, USA

3-Physical and Life Sciences Directorate, Lawrence Livermore National Laboratory, Livermore, California 94550, United States

4-Department of Physics, Washington University in Saint Louis, Saint Louis, USA

5-Research Center for Functional Materials, National Institute for Materials Science, 1-1 Namiki, Tsukuba, Japan 8

6-International Center for Materials Nanoarchitectonics, National Institute for Materials Science, 1-1 Namiki, Tsukuba,Japan

7 Material Science & Technology Division, Oak Ridge National Laboratory, Oak Ridge, Tennessee 37831, USA

ABSTRACT

We investigate the electronic properties of a graphene and α-ruthenium trichloride (α-RuCl$_3$) heterostructure, using a combination of experimental and theoretical techniques. α-RuCl$_3$ is a Mott insulator and a Kitaev material, and its combination with graphene has gained increasing attention due to its potential applicability in novel electronic and optoelectronic devices. By using a combination of spatially resolved photoemission spectroscopy, low energy electron microscopy, and density functional theory (DFT) calculations we are able to provide a first direct visualization of the massive charge transfer from graphene to α-RuCl$_3$, which can modify the electronic properties of both materials, leading to novel electronic phenomena at their interface. The electronic band structure is compared to DFT calculations that confirm the occurrence of a Mott transition for α-RuCl$_3$. Finally, a measurement of spatially resolved work function allows for a direct estimate of the interface dipole between graphene and α-RuCl$_3$. The strong coupling between graphene and α-RuCl$_3$ could lead to new ways of manipulating electronic properties of two-dimensional lateral heterojunction. Understanding the electronic properties of this structure is pivotal for designing next generation low-power opto-electronics devices.

# INTRODUCTION

In recent years, there has been a surge in interest in heterostructures composed of different two-dimensional (2D) materials[1–3]. These systems offer unique electronic properties that arise from their interfacial interactions, making them promising candidates for novel electronic and optoelectronic devices[4,5]. The absence of a covalent chemical bond between the layers opens the route toward designing 2D quantum systems that hold the promise to unlock the post-Moore era[6,7]. One particularly exciting development is the recent discovery of a permanent charge transfer induced in graphene by proximity to α-RuCl₃ (RuCl₃ hereafter). In turn, this offers a route to exploring the physics of charge-doped Mott insulators[8–12].

RuCl₃ is a layered transition metal compound with a honeycomb lattice structure similar to graphene. However, unlike graphene, it is a Mott material with insulating behaviour arising from strong electronic correlations[13]. At low temperatures, the complex competition of magnetic interactions ultimately stabilizes a zigzag antiferromagnet in RuCl₃[14]. However, the influence of doping, for instance by photoinduced charged carriers is predicted to stabilize ferromagnetic order[15]. Additionally, RuCl₃ is classified as a Kitaev material due to its strong spin-orbit coupling, crystal field, and electronic correlations, which lead to anisotropic exchange interactions that favour the formation of a quantum spin liquid expected to host Majorana fermions[16–20]. These quasiparticles have non-Abelian statistics and are essential for topological quantum computation[21]. Nonetheless, the evidence for such behaviour in RuCl₃ is still under debate[22] and seem to be strongly affected by the presence of crystal defects which promote impurity scattering and non-Kitaev interactions[23].

When graphene is brought in contact with RuCl₃, a charge transfer occurs between the two materials due to their different work functions and electronic structures[8]. This charge transfer can modify and hybridize the electronic properties of both materials[10], as well as influence the magnetism in RuCl₃[15,24]. The coupling between graphene and RuCl₃ can modify the electronic band structure of RuCl₃ and enhance its spin-orbit coupling, potentially impacting the Kitaev physics in the material[9,24]. Anomalous quantum oscillations have been reported in the Gr/RuCl₃ heterostructure and explained within the Kitaev-Kondo lattice model[25–27]. The strong charge transfer has also been used to create modulation-doped graphene where a lateral thickness variation of a tunnel barrier changes the magnitude of the charge transfer between graphene and RuCl₃[11], enabling ultra-sharp (less than 5-nm-wide) p-n junctions[12], which were also observed in nanobubbles of graphene on RuCl₃[28]. The interaction between graphene and RuCl₃ can also lead to plasmon polaritons at the interface[29]. The coupling between plasmon polaritons and the Mott physics in RuCl₃ could unlock new ways of manipulating light and electronic properties, with potential applications in sensing, imaging, and communication. Moreover, by leveraging the unique passive doping control (no gating needed) of RuCl₃ over graphene, we envision the creation of low-power devices that exhibit enhanced light harvesting capabilities and precise control over optical signals[30].

Here, we employ a combination of experimental and theoretical techniques to better investigate the electronic properties of the interface between RuCl₃ and graphene. Nanometer-scale spatially resolved photoemission spectroscopy (nanoXPS), low energy electron microscopy (LEEM), and density functional theory (DFT) calculations are used to explore the electronic properties of the heterostructure allowing for a first direct visualization of its charge transfer.

Elemental core levels of 2D systems can be mapped effectively via nanoXPS, and the dispersive electronic band structure of the heterostructure with submicron spatial resolution via angle resolved photoemission spectroscopy (nanoARPES)[31]. LEEM allows for imaging the morphology and electronic properties of heterostructures with high spatial resolution. By using low energy electrons to probe the surface of the material, local variations in electronic properties can be investigated to study their evolution over time. DFT calculations provide complementary insights into the electronic properties of the heterostructure, allowing us to simulate the electronic structure of the material. By comparing our experimental results with the DFT calculations, we validate our findings and provide a more complete understanding of the electronic properties of the heterostructure.

The experimental data show a massive charge transfer from graphene to RuCl₃, clearly visible in the nanoARPES data and reflected in core levels measured via nanoXPS. LEEM analysis provides a value of the shift in work function that is much lower than the band shift measured via nanoARPES, consequent to the charge transfer between the layers. This discrepancy can be attributed to the presence of a dipole at the interface that greatly affects the work function value[32]. Moreover, a Mott transition is induced and a band close to the Fermi level is observed. Our DFT calculations suggest that this effect is caused by the extra carriers injected in RuCl₃.

**RESULTS AND DISCUSSION**:

We fabricated a heterostructure composed of a thick hexagonal boron nitride (h-BN) substrate, with three other materials exfoliated on top: graphene, thin h-BN (2nm), and RuCl₃ (Fig. 1(a)). The fabrication and experimental details are reported in the supporting information. The device has three distinct regions: one with graphene on thick h-BN as a reference, one with graphene directly on RuCl₃, and one with a thin h-BN flake sandwiched between graphene and RuCl₃. The thin h-BN acts as a buffer to decrease the RuCl₃ /graphene interactions[28]. The thick h-BN substrate provides a stable and flat surface for other materials and minimizes the effect of the underlying substrate on the electronic properties of the heterostructure. A sketch of the three regions is reported in Fig. 1(b) with coherent color scheme. Fig. 1(c) displays the heterostructure contour, where the contrast is given by the counts of the photoelectrons coming from the valence band of RuCl₃ at binding energy -1.3 eV. The contrast allows for identifying the three regions described above. The color scheme for the three colored squares on the map is consistent with the sketch in panel (b) and confirmed by the core level analysis via XPS. We focus on the peaks originating from Cl, Ru, and C core levels, reported in Fig. 1(d) and Fig. 1(e). The most informative peak regarding the location of the RuCl$_3$ region is the Cl 2p core level. Its signal decreases when the h-BN buffer is present and disappears entirely outside the RuCl$_3$ flake. The Ru 3d$_{3/2}$ core level partially overlaps with the C 1s. It is possible to fit and track the evolution of the C 1s peak, for the three different regions (Fig. 1 (f)). The fitting is performed considering one Doniach Sunjic (DS) asymmetric lineshape[33] for graphene and one DS for Ru 3d$_{3/2}$, plus a gaussian peak to take into account the broad and weak contribution from the 3d$_{5/2}$ peak. While the Ru level remains roughly at fixed position, the C peak progressively shifts towards lower binding energy when increasing the coupling strength between Gr and RuCl$_3$. An overall shift of about -750 meV is observed for C 1s from isolated graphene. RuCl$_3$ is expected to induce a significant electron depletion in graphene[8,9] that is reflected on an electrostatic shift of the C core levels and the whole graphene band structure.

To directly visualize the electronic properties of the system, we conducted a nanoARPES study. This study enabled us to observe the electronic band structure in three specific regions, highlighted in Fig. 1(c). In Fig. 2(a), we present the bands of the graphene-only region, which are close to the neutrality point. In Fig. 2(b) and (c), we show the band structures of the RuCl$_3$/h-BN/Gr and RuCl$_3$/Gr regions, respectively. Notably, the graphene bands in these regions are shifted upwards by approximately 500 meV where the h-BN layers separate the Gr and RuCl$_3$ crystals, and by 750 meV where RuCl$_3$ is in direct contact with graphene, indicating a progressive p-doping when reducing the distance between graphene and RuCl$_3$. The effect of the charge transfer and the consequent interface dipole is reflected in the 1 eV upward shift of the h-BN bands (white arrows) relative to the region lacking RuCl$_3$ reported in Fig. 2(a). Because of the short mean free path of the photoelectrons, the RuCl$_3$ bands are only visible in the region where graphene is in direct contact, with no intervening h-BN buffer. The RuCl$_3$ bands are identified by studying the energy distribution curves (EDCs) taken along the dashed line in panels (a-c) of Fig. 2 (Fig. 2(d)). The RuCl$_3$ electronic structure, highlighted by three white arrows, displays two dispersionless bands centered at binding energy ~0.5 and ~1.3 eV, and a third more dispersive band at a deeper binding energy (~3,8 eV).

A clearer estimate of the total amount of charge transferred between the layers, with and without the h-BN buffer layer, is given by considering the Fermi surface for each of the three regions, as reported in Fig. 2(e-g). The momentum distribution curves (MDCs) collected along the dashed lines on the Fermi surface are displayed in Fig. 2(h). By fitting with two Lorentzian curves, the position of their maxima is used to evaluate the Fermi surface area, approximated as a circle. By means of the Luttinger theorem[34–36] we can extract the amount of charge tunneling from graphene to RuCl$_3$, about $4.1 \times 10^{13}$ cm$^{-2}$, consistent with previous, if indirect, experimental and computational observations[8–10,24,29]. When the spacing between layers is increased with a few h-BN layers, the tunnel barrier thickens, resulting in a decreased charge transfer and therefore a lower p-doping level in graphene (~ $1.7 \times 10^{13}$ cm$^{-2}$).

To interpret the measured band structure, we performed DFT calculations with a RuCl$_3$ supercell both with and without graphene as described in the supporting information. The nanoARPES experiment was performed at room temperature, indicating that the RuCl$_3$/Gr heterostructure was likely in a paramagnetic configuration. However, a proper representation of a paramagnetic state in a DFT simulation would require a much larger supercell resulting in prohibitive computational workloads. To better understand the effect of graphene on RuCl$_3$ bands, we used a simpler model that allowed for an antiferromagnetic structure with and without graphene. For sake of simplicity, and since magnetic ordering in RuCl$_3$ mostly occurs along one of the in-plane directions, we considered spin resolved collinear magnetism.

Fig. 3(a) displays the calculated bands of the pristine RuCl$_3$ structure, reflecting the highly correlated nature of the system with non-dispersive bands and a gap opening around the Fermi level. The gap presents only a qualitative agreement due to the choice of the Hubbard parameters and the collinear magnetism, since a more accurate description would require a non-collinear spin approach that takes into account spin-orbit coupling too. However, we were able to quantify the band structure changes caused by the charge doping, by comparing the electronic structures of the pristine RuCl$_3$ and graphene-doped heterostructure supercells within the same collinear spin polarized approach. Fig. 3(b) displays the bands of the RuCl$_3$/Gr supercell, that agree well with the observed shift of the neutrality point by about 750 meV in Fig. 2(c).

In the nanoARPES data we observed the presence of a broad band with spectral weight centered at about 0.5 eV below the Fermi level, which has not reported in prior ARPES experiments on bulk RuCl$_3$[37,38]. In Fig. 3(c), we compare the pristine RuCl$_3$ bands with those of the RuCl$_3$/Gr system. It is immediately apparent that in the heterostructure the presence of graphene pushes an unoccupied band of RuCl$_3$ below the Fermi level, populating the previously gapped region of the spectra. It has been demonstrated that the presence of dopants on the RuCl$_3$ surface causes new bands to be populated near the Fermi level, attributed to a Mott transition where the system retains spin correlations. We therefore infer that RuCl$_3$ undergoes a similar Mott transition to the one observed with Rb and K atoms in Ref. [37].

Finally, we can quantify the electric dipole generated at the interface by the charge transfer to the RuCl$_3$. It is possible to compute the magnitude of the dipole by measuring the variation of the work function across the different regions of the system. By applying a positive voltage to the sample, the incident LEEM electrons transition from mirror mode, with the electrons reflecting before touching the sample surface, to LEEM mode, where the electrons are scattered from sample surface with a landing energy proportional to the applied sample bias. In LEEM mode the incident electrons can be accepted into unoccupied bands of the sample surface causing a lower reflected intensity than in the mirror mode. The inflection point of this drop in intensity from mirror mode to LEEM mode can be interpreted as the work function of the sample surface when accounting for the work function of the LEEM cathode.

In Fig. 4 (a) selected LEEM images collected just below the mirror mode transition display the boundary of the three regions discussed above. The line profiles in Fig. 4 (b-d) show the average work function across each boundary in the directions indicated by the arrows in the corresponding LEEM images. The profile analysis highlights a difference in the work function of about 230 meV across the interface between RuCl₃/Gr and Gr. When the h-BN buffer layer is also present, the shift in the work function is reduced by 160 meV. This difference with respect to the graphene region agrees with the 70 meV difference between the RuCl₃/h-BN/Gr and RuCl₃/Gr regions. Previously, Yu and co-workers demonstrated that the work function of graphene can be substantially affected by the dipole formed by surface adsorbates[32]. Analogously, here we estimate the magnitude of the electric field at the interface knowing the value of the chemical potential, and the value of the work function, with respect to pristine graphene. In the presence of a dipole at the interface, the work function of graphene can be written as

$$W_{sample} = \Delta W_D + W_{gr}^0 - E_F, \qquad \text{Eq (1)}$$

where $\Delta W_D$ is the offset of work function due to the dipole at the interface, $W_{gr}^0$ is the intrinsic work function of the undoped graphene and $E_F$ is the position of the Fermi level[32]. We can therefore evaluate the magnitude of the electric dipole with respect to the pristine sample simply considering the measured work function difference across the different regions and adding this to the corresponding difference in $E_F$ position with respect to the Dirac point. This results in an electric dipole energy of ~ 1 eV and ~ 660 meV for RuCl$_3$/Gr and RuCl$_3$/h-BN/Gr structures, respectively.

In conclusion, we used a combination of experimental techniques, including nanoXPS and nanoARPES, LEEM, and DFT calculations to investigate the electronic properties of the RuCl$_3$/Gr heterostructure. The results showed direct evidence of significant charge transfer from graphene to RuCl$_3$, leading to a doping-induced Mott transition and potential enhancement of Kitaev physics. LEEM measurement also allowed us to provide an estimate of the dipole moment formed at the interface between RuCl$_3$ and graphene, instrumental for a comprehensive device modeling. This work lays out valuable insights into the electronic properties of RuCl$_3$/Gr heterostructures and its potential for future applications where the passive control of the doping level in graphene is at the foundation of low-power electronics and light-harvesting devices.

**Author Contributions**

The manuscript was written through contributions of all authors. All authors have given approval to the final version of the manuscript.

**Corresponding Author**

Antonio Rossi: antonio.rossi@iit.it

Erik A. Henriksen: henriksen@wustl.edu

**Present Addresses**

Antonio Rossi: Center for Nanotechnology Innovation @ NEST, Instituto Italiano di Tecnologia, Pisa, 56127 Italy

Jesse Balgley: Department of Mechanical Engineering, Columbia University, New York, NY 10027, USA

**Acknowledgements**

The authors thank Dr. Luca Moreschini from University of California, Berkeley for the fruitful discussion.

This work was supported as part of the Center for Novel Pathways to Quantum Coherence in Materials, an Energy Frontier Research Center funded by the U.S. Department of Energy, Office of Science, Basic Energy Sciences. Work was performed at the Molecular Foundry and at the Advanced Light Source, which was supported by the Office of Science, Office of Basic Energy Sciences, of the U.S.Department of Energy under contract no. DE-AC02-05CH11231. R.D. performed this work under the auspices of the U.S. Department of Energy by Lawrence Livermore National Laboratory under Contract DE-AC52-07NA27344. EAH acknowledges support by the Office of the Under Secretary of Defense for Research and Engineering under

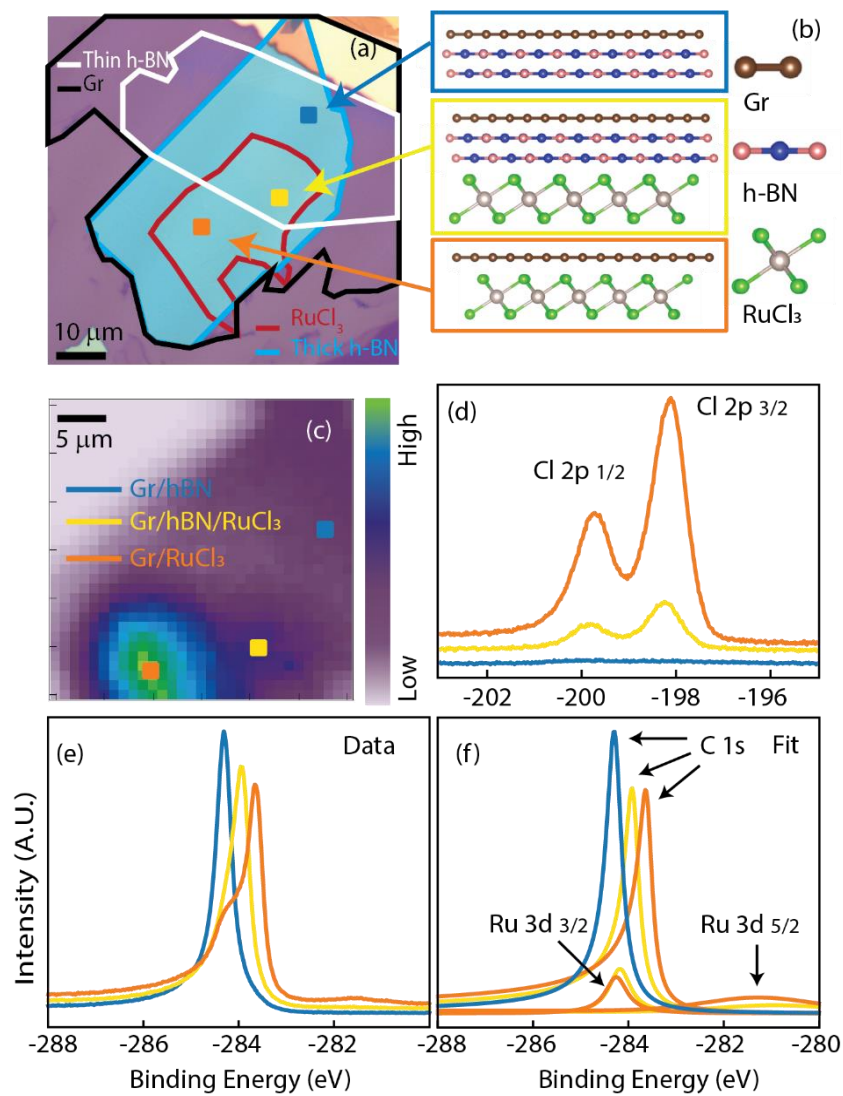

*Figure 1 (a) Optical image of the analysed device. False-color contours are used to highlight the different layers. (b) Sketch of the side view of three regions of interest. RuCl$_3$ must be considered as multilayers, even though one layer is depicted for neatness. (c) Photoelectron intensity map collected at $E-E_F = -1.3$ eV. (d) Cl 2p core level collected in the regions highlighted in panel (c) with the corresponding color scheme. (e) Ru 3d and C 1s core levels collected from the same points of panel (d). (f) Fit of the Ru 3d and C 1s level for the data reported in panel (e).*

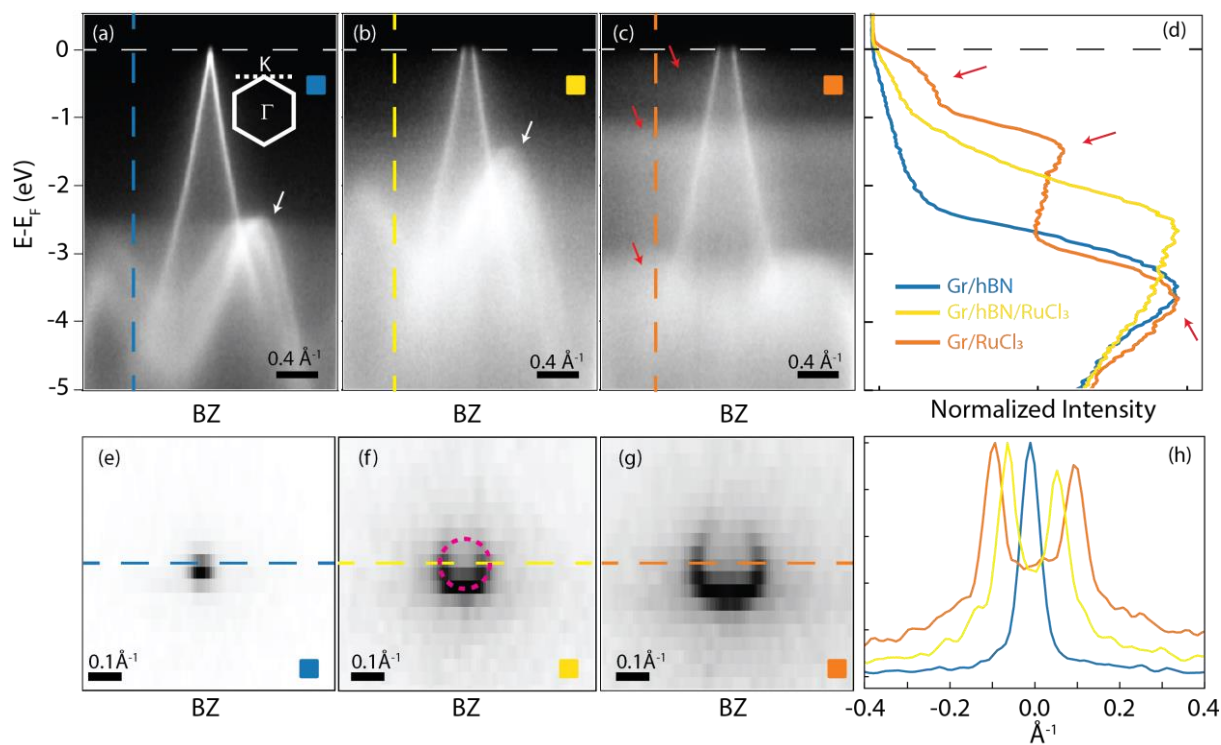

*Figure 2 (a-c) Band structure collected around graphene K (point as depicted by panel (a) inset) from the three regions described above with consistent color scheme. The horizontal dashed line is the Fermi level. The white arrows highlight the h-BN bands, the red arrows the $RuCl_3$ bands. (d) EDCs collected from panels (a-c) along the vertical dashed line. The red arrows highlight the corresponding states in panel (c). (e-g) Fermi surface of the band structure from the sample regions with the corresponding color scheme. The dashed red circle in panel (f) approximates the graphene Fermi surface. (h) MDCs collected across the dashed lines in panels (e-g)*

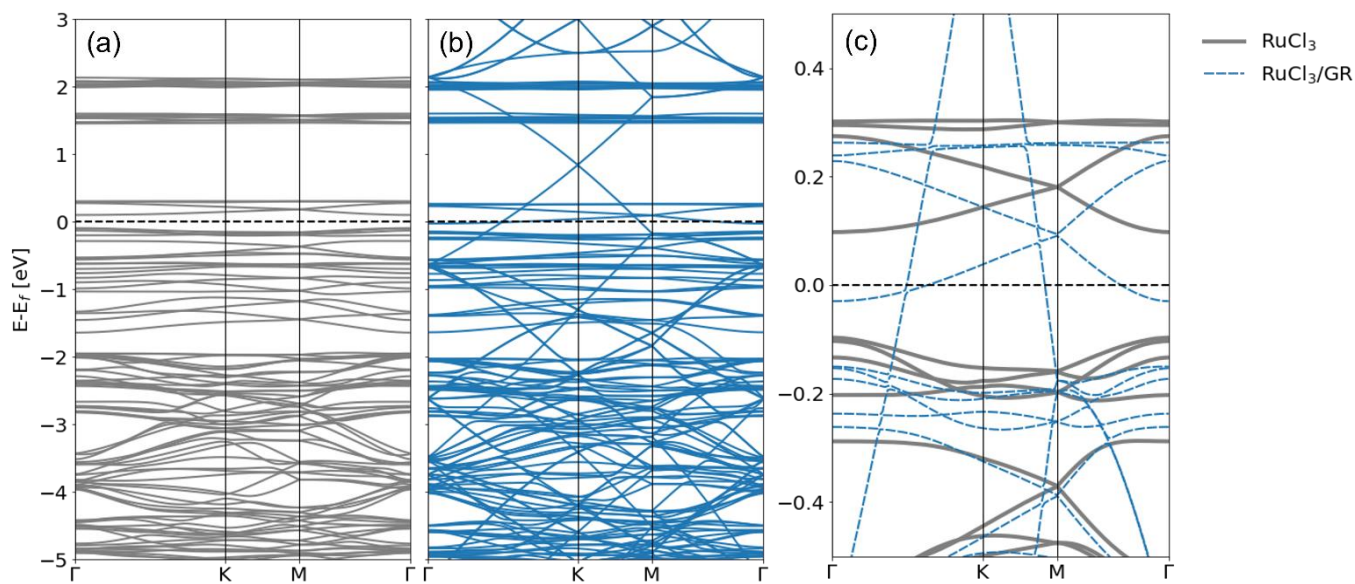

*Figure 3 (a) Band structure for a 2x2 antiferromagnetic zig-zag $RuCl_3$ supercell, (b) band structure for an antiferromagnetic zig-zag $RuCl_3$/gr heterostructure, (c) band structure comparison between (a) and (b) in the interval (-0.5,0.5) eV.*

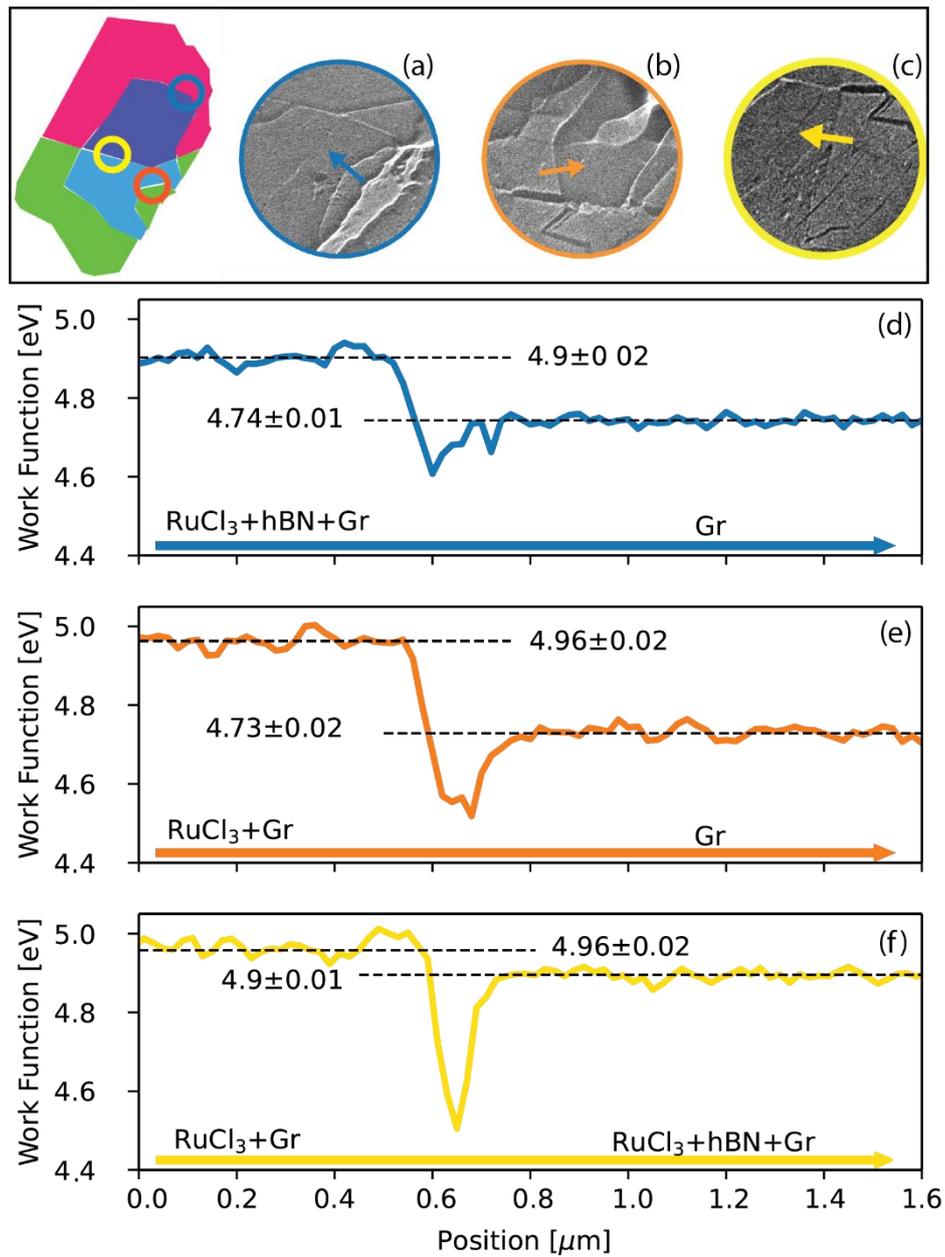

Figure 4 (a-c) LEEM image of selected regions of interest (a) RuCl3-hBN-Gr / Gr, (b) RuCl3-Gr/Gr, (c) RuCl3-Gr / RuCl3-hBN-Gr. (d-f) Intensity profile across the boundary separating the different regions along the arrows drawn in the microscopy panels.

# Supporting Information: Direct visualization of the charge transfer in Graphene/α-RuCl$_3$ heterostructures.

*Antonio Rossi[1,2]\*, Riccardo Dettori[3], Cameron Johnson[2], Jesse Balgley[4], John C. Thomas[2], Luca Francaviglia[2], Andreas K. Schmid[2], Kenji Watanabe[5], Takashi Taniguchi[6], Matthew Cothrine[7], David G. Mandrus[7], Chris Jozwiak[1], Aaron Bostwick[1], Erik A. Henriksen[4], Alexander Weber-Bargioni[2] and Eli Rotenberg[1]*

1-The Molecular Foundry, Lawrence Berkeley National Laboratory, Berkeley, USA.

2-Advanced Light Source, Lawrence Berkeley National Laboratory, Berkeley, USA

3-Physical and Life Sciences Directorate, Lawrence Livermore National Laboratory, Livermore, California 94550, United States

4-Department of Physics, Washington University in Saint Louis, Saint Louis, USA

5-Research Center for Functional Materials, National Institute for Materials Science, 1-1 Namiki, Tsukuba, Japan 8

6-International Center for Materials Nanoarchitectonics, National Institute for Materials Science, 1-1 Namiki, Tsukuba,Japan

7 Material Science & Technology Division, Oak Ridge National Laboratory, Oak Ridge, Tennessee 37831, USA

**Device fabrication**

Graphene, hexagonal boron nitride (hBN), and RuCl₃ flakes were isolated via mechanical exfoliation and atomic force microscopy was used to confirm the flake thicknesses (Fig. S1). A thick h-BN flake is used as a substrate for the device ensuring a flat support surface. The sample was prepared using a dry van der Waals stacking technique[39] to pick up flakes of graphene, thin hBN (~2 nm), RuCl₃, and thick hBN (~40 nm), successively, using an adhesive layer of poly(bisphenol A carbonate) (PC). The heterostack was deposited at 180 °C onto a prepatterned gold pad to ensure a contact with the ground, which was thermally evaporated on an SiO$_2$/p-Si substrate.

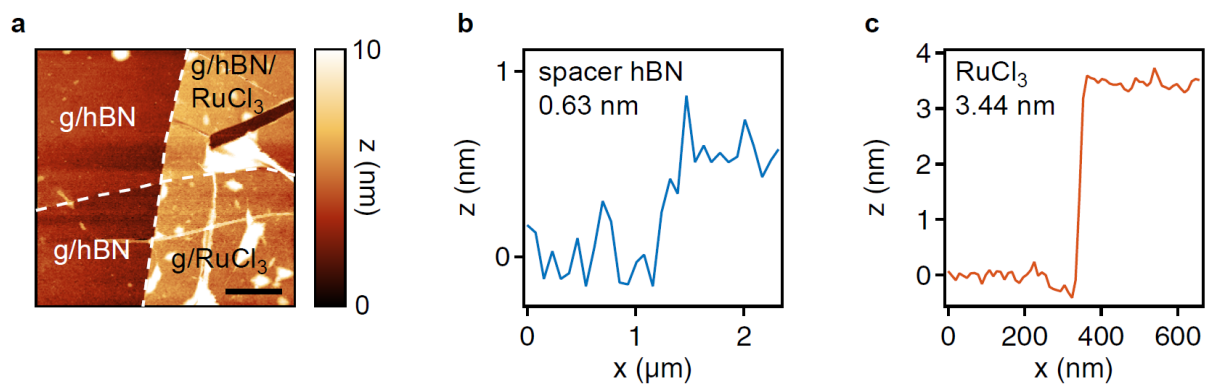

Fig S 1 (a) AFM image of the three regions of interest, scale bar is 500 nm. (b) AFM profile characterization of the h-BN layer sandwiched between graphene and RuCl₃. (c) AFM profile characterization of the RuCl₃ flake.

Polymer residues are removed by nanobrooming with an atomic force microscope [40]. We first acquired images to identify the area to be cleaned. We used a Park NX10 microscope in non-contact AFM mode with PPP-NCHR probes manufactured by Nanosensors (nominal resonance of 330 kHz, nominal stiffness of 42 Nm$^{-1}$). We then switched to contact mode for the actual cleaning and scanned the selected area with Arrow-CONTPt probes by NanoWorld

(nominal resonance 14 kHz, nominal force constant 0.2 Nm/m). We typically chose values between 65 nN and 100 nN as force setpoints in contact mode and scan rates of 0.5-1 Hz. We used dull tips and assumed a tip radius of about 5nm. We therefore set the distance between two adjacent scanning lines to be equal or shorter than 5nm not to leave dirty gaps.

**Experimental details**

NanoARPES measurements are performed at beamline 7.0.2 at the Advanced Light Source (ALS) facility at Lawrence Berkeley National Laboratory. The lateral resolution is below 1 um. The nanoARPES data are collected using 150 eV photon energy, while 350 eV excitation is used for core level analysis (nanoXPS). Linear-horizontal polarized light is employed. All the measurements were performed at room temperature.

LEEM measurements were performed on the quantum spin polarized LEEM in the Molecular Foundry at Lawrence Berkeley National Laboratory. LEEM is a UHV electron microscope that uses electron optics to prepare a plane wave of electrons which reflect/scatter from the sample surface with landing energies ranging from 0-500 eV. The reflected/scattered electrons are then collected by imaging optics to project a real or momentum space image of the surface onto a CMOS electron camera.

DFT simulations were performed with the VASP ab initio simulation package[41–43], using the projector augmented wave (PAW) method[44,45] and the Perdew, Burke, and Ernzerhof exchange-correlation functional (PBE)[46]. Partial occupancies of the electronic states were set with fourth order Methfessel-Paxton smearing[47], using a width of 0.2 eV. We observed converged energies for the bulk system with a plane-wave energy cutoff of 500 eV, with a self-consistent field (SCF) convergence criterion of $10^{-6}$ eV. The force convergence tolerance was set to 0.01 eV/Å for each atom in each direction. During optimization, atomic positions were relaxed using a 5 × 5 × 1 k-point Monkhorst-Pack mesh sampling of Brillouin zone[48], where the smallest mesh value is related to the nonperiodic direction of the system. Due to the lattice mismatch between α-RuCl₃ and graphene (2.46 Å), the RuCl₃+Gr heterostructure was realized as a supercell: in particular, we realized a hexagonal supercell containing 82 atoms (8 Ru, 24 Cl, 50 C) composed of a 5x5 graphene sheet and a fully relaxed 2x2 RuCl₃ monolayer (similarly to ref.[9] ). This setup resulted in a lattice mismatch lower than 2%, thus minimizing the strain on the RuCl3 layer. To avoid spurious interactions with periodic replicas in the out-of-plane direction the simulation cell consisted in a 20 Å thick vacuum region along the z-axis.

In order to account for electronic correlations more accurately, we have adopted the rotationally invariant approach to GGA+U of ref [49]: the on-site coulomb (U) and exchange (J) corrections were applied to Ru *d*-orbitals. In agreement with the literature, and after performing extensive tests, we proceeded running our calculations with $U_{eff}$=2.0 eV (U=2.5 and J=0.5 eV)[15,50,51].